\documentclass[aps,prl,reprint,superscriptaddress,nofootinbib,onecolumn,11pt]{revtex4-2}

\usepackage{comment}
\usepackage[utf8]{inputenc} 
\usepackage{graphicx,color,overpic,mathtools}
\usepackage{amsthm,amsmath,amssymb,hyperref,mathrsfs}
\usepackage{braket,bm,bbm,setspace}
\PassOptionsToPackage{normalem}{ulem}
\usepackage{ulem} 
\usepackage{physics}
\usepackage{float}
\usepackage[makeroom]{cancel}
\usepackage[english]{babel}
\usepackage{graphicx}
\graphicspath{ {images/} }
\addto\captionsspanish{}
\hypersetup{
    colorlinks=false,
    pdfborder={0 0 0},
}

\begin{document}

\title{Semiclassical relativistic stars}

\author{Julio Arrechea}
\author{Carlos Barcel\'o} 
\affiliation{Instituto de Astrof\'isica de Andaluc\'ia (IAA-CSIC),
Glorieta de la Astronom\'ia, 18008 Granada, Spain}
\author{Ra\'ul Carballo-Rubio}
\affiliation{CP3-Origins, University of Southern Denmark, Campusvej 55, DK-5230 Odense M, Denmark}
\affiliation{Florida Space Institute, University of Central Florida, 12354 Research Parkway, Partnership 1, 32826 Orlando, FL, USA}
\author{Luis J. Garay} 
\affiliation{Departamento de F\'{\i}sica Te\'orica and IPARCOS, Universidad Complutense de Madrid, 28040 Madrid, Spain}  
\affiliation{Instituto de Estructura de la Materia (IEM-CSIC), Serrano 121, 28006 Madrid, Spain}

\begin{abstract}
We present strong evidence that semiclassical gravity can give place to self-consistent ultracompact stars beyond the Buchdahl limit. We integrate the semiclassical equations of (spherically symmetric) stellar equilibrium for a constant-density classical fluid. The semiclassical contribution is modelled by a quantum massless scalar field in the only static vacuum state compatible with asymptotic flatness (Boulware vacuum). The Renormalized Stress-Energy Tensor (RSET) is firstly approximated by the analytic Polyakov approximation. This  already reveals a crucial difference with respect to purely classical solutions: stars with compactness close to that of a black hole exhibit bounded pressures and curvatures up to a very small central core compared with the star radius. This suggests that a more refined approximation to the RSET at the core may give rise to strictly regular configurations. Following this suggestion, we prove that a minimal deformation of the Polyakov approximation inside the central core is sufficient to produce regular ultracompact stellar configurations.
\end{abstract}

\maketitle

\section*{Introduction}

In the dawn of   gravitational-wave astronomy, we are closer than ever to unveiling the mystery that surrounds astrophysical black holes. One of the main outstanding questions is whether these astronomical objects lurking in the universe are General Relativity (GR) black holes---in the sense of having long-lived horizons and essentially empty interiors---or material stellar-like objects with no horizons. There exists no conclusive experimental confirmation that astrophysical black holes must correspond to strictly classical (or Hawking-evaporating) black holes~\cite{Abramowiczetal2002, Eckartetal2017, CardosoPani2017, Carballo-Rubioetal2018, CardosoPani2019}. On the other hand, the standard paradigm of black hole evaporation is not free from problems, which have been discussed for more than 40 years \cite{Hawking1976, Page1993, UnruhWald2017, Barceloetal2015}. Given the new observational possibilities, the search for viable models of exotic compact-and-dark objects as alternatives to GR black holes is becoming popular, if only as a catalog with which to compare the GR predictions~\cite{CardosoPani2019}. Although probing them observationally is not easy, it is also far from hopeless~\cite{AbediAfshordi2016, Barceloetal2016, UrbanoVeermae2018, Raposoetal2018,  Giddings2019, Maggioetal2020, Zulianelloetal2020, Ikedaetal2021}.

The current alternative models   involve more or less exotic new physics according to taste. Many of them investigate new forms of stellar equilibrium which could represent intermediate steps in the stellar ladder between neutron stars and GR black holes (e.g. boson stars~\cite{Kaup1968,SchunckMielke2003}, fluid stars~\cite{ShapiroTeukolsky1983,Saidaetal2015}, anisotropic stars~\cite{Andreasson2008,Raposoetal2019}, and solutions obtained through minimal and complete geometric deformation approaches \cite{Ovalle2017,Mauryaetal2021}).  Other models propose that GR black holes are only a mathematical approximation to a more complex situation in which, from a few Planck lengths above the gravitational radius (where the event horizon would appear in GR) inwards, the spacetime geometry is substantially different or even ceases to exist. Examples in which the internal spacetime geometry is substantially modified include gravastars~\cite{MazurMottola2004,MazurMottola2015} and, closer to the spirit of our approach, the 2+2 hole geometries~\cite{HoldomRen2017} based on a modified gravity prescription. Examples in which the classical notion of spacetime ceases to exist in the interior include fuzzballs~\cite{Mathur2005} and collapsed polymers~\cite{BrusteinMedved2017}. 

Our own approach is to search for semiclassical relativistic stars. Semiclassical gravity amounts to a straightforward modification of classical GR that takes into account vacuum polarization effects, analogous to those observed for instance in Quantum Electrodynamics. Thus, it is a conceptually simple and rather conservative framework.
This approach suggests that trapping horizons are possible only as a transient notion, never to be found as a static property~\cite{Berthiereetal2017, Arrecheaetal2020, Arrecheaetal2021}.
Within semiclassical gravity we will provide the strongest theoretical evidence thus far of the existence of stellar configurations beyond relativistic stars.

\section*{Semiclassical gravity}

It has its roots in the idea that spacetime curvature must locally deform the energetic contribution of the zero-point fluctuations of quantum fields in a way that cannot be renormalized away. It assumes the preservation of an effective classical spacetime structure, introducing only an additional zero-point Stress-Energy-Tensor (SET)   $T^{\rm ZP}_{\mu\nu}$ into the Einstein equations: 
\begin{equation}\label{Eq:SemiEinstein}
G_{\mu\nu}=8\pi \left(T_{\mu\nu}+ T^{\rm ZP}_{\mu\nu} \right).
\end{equation}
This definition deliberately omits leaning towards any specific operational procedure followed to obtain the $T^{\rm ZP}_{\mu\nu}$.

The method usually followed to find   physically sensible expressions for $T^{\rm ZP}_{\mu\nu}$ consists in promoting the corresponding SET into an operator in the quantum field theory (QFT) whose expectation value can be computed in a suitably chosen vacuum state: $T^{\rm ZP}_{\mu\nu}=\langle\hat{T}_{\mu\nu}\rangle$~\cite{BirrellDavies1982, HuVerdaguer2020, Simon1991, FlanaganWald1996}. As the SET operator is not   well defined in the quantum theory~\cite{Wald1978}, this procedure requires   regularization and renormalization. The resulting Renormalized SETs (RSETs)
have several shortcomings. First, the outcome is not unique, exhibiting ambiguities~\cite{Wald1978}. 
Second, in generic situations in 3+1 dimensions the resulting RSETs have higher-derivative terms \cite{ParkerSimon1993, Simon1991, Andersonetal1993, Fabbrietal2003}, which hinder the search of reliable self-consistent solutions. 
Additionally, these RSETs might even lack
a closed analytic form \cite{Andersonetal1995, Grovesetal2002, Levietal2016, Andersonetal2020}.  
 
Even with these shortcomings in mind, it is important to realize that semiclassical effects have some robust   generic features. (i) The RSETs can and must provide violations of the pointwise energy conditions \cite{Visser1996, BarceloVisser2002, Fewster2012, Curiel2017}, if only to be able to encode effects like Hawking evaporation.
This evaporative process is caused by a combination of negative and positive matter fluxes that penetrate the horizon and escape to infinity, respectively. In the static situations considered here, the RSET accounts for (the most part negative) energy and pressure contributions coming from the vacuum state of the quantum scalar field and that permeate the entire spacetime.
This by itself indicates the potentiality of avoiding standard classical results such as the singularity theorems \cite{Hawking1970}. (ii) Although the semiclassical corrections appear multiplied by the Planck constant, there are scenarios---essentially when matter   remains extremely close to its gravitational radius---in which energy-condition violations can become huge~\cite{Barceloetal2009, Visseretal2009, BanerjeeParanjape2009, Barceloetal2019}.
(iii) The RSET naturally brings  anisotropic pressures into GR even if absent at the classical level. (iv) As we will show in this work, the total energy density (i.e. the sum of classical and semiclassical contributions) can decrease inwards in some layers of a stellar structure even when the classical densities are compelled to grow inwards.  
These generic features suggest the possibility of violating the Buchdahl compactness limit for classical stars~\cite{Buchdahl1959, UrbanoVeermae2018, CardosoPani2019}.

Finally, let us stress the remarkable fact that semiclassical gravity effects cannot coexist with static horizons of any kind~\cite{Berthiereetal2017, Arrecheaetal2020, Arrecheaetal2021}. Then, within semiclassical gravity one has either objects with evaporating horizons (the standard view), or genuinely static configurations without horizons. 

\section*{Beyond the Polyakov approximation}

We analyze the semiclassical Eqs. \eqref{Eq:SemiEinstein} seeking to find the form of spherically symmetric, static, and asymptotically flat self-consistent solutions, i.e., geometries that can represent stellar objects. The corresponding line element is
\begin{equation}\label{Eq:LineElement}
    ds^{2}=-e^{2\phi(r)}dt^{2}+\left[1-C(r)\right]^{-1}dr^{2}+r^{2}d\Omega^{2},
\end{equation}
where $d\Omega^{2}$ is the angular line element of the 2-sphere. The functions $e^{2\phi}$ and $C$ denote the redshift and the compactness, respectively. The former measures the   redshift suffered by outgoing null rays \cite{Hayward1993} and the latter is the quotient $2m(r)/r$, with $m(r)$ the Misner-Sharp mass \cite{MisnerSharp1964, HernandezMisner1966, Hayward1994}.

To   explore the   characteristics of the set of semiclassical solutions, we consider a regularized version of the Polyakov RSET of a single massless scalar field~\cite{Polyakov1981,DaviesFulling1977,Arrecheaetal2020}. We require the vacuum state to be   Boulware's---the only vacuum   consistent with static and asymptotically flat stellar-like solutions \cite{BirrellDavies1982}---. The Polyakov RSET is an approximation to the zero-point SET which is based on (i) modeling the propagation as if it   effectively happened in a reduced $1+1$ spacetime (the $t,r$ sector of the metric) and (ii) neglecting backscattering   due to the gravitational potential~\cite{FabbriNavarro-Salas2005}. However crude these two simplifications may be, they bring in exchange an RSET which is unique, analytic, properly captures the defining features of vacuum states \cite{Hiscock1977b}, and contains only up to second derivatives of the metric, which allows to define a differential problem equivalent to that of classical GR. We then adopt a modified-gravity philosophy and find its associated solutions.

The Polyakov approximation leads to the RSET  
\begin{align}
&  \langle \hat{T}_{rr} \rangle=
    -F~\frac{l_{\rm P}^{2} \psi^{2}}{8\pi},
\nonumber \\
&  \langle\hat{T}_{tt}\rangle= 
    F~\frac{l_{\rm P}^{2}e^{2\phi}}{8\pi}\left[2\psi'\left(1-C\right)+\psi^{2}\left(1-C\right)-\psi C'\right],
\nonumber \\
& \langle \hat{T}_{\theta\theta} \rangle=\frac{\langle \hat{T}_{\varphi\varphi}\rangle}{\sin^{2}\theta}=    
-\left(2F+r F'\right)~\frac{l_{\rm P}^{2}r^{2}}{16\pi}\left(1-C\right)\psi^{2},
\end{align}
where $\psi=\phi'$ (the $'$ denoting derivatives with respect to $r$) and $F$ a radial function. For the Polyakov RSET,
\begin{equation}\label{eq:4dimRSET}
F(r)=\frac{1}{r^2}.
\end{equation}
The divergence of \eqref{eq:4dimRSET} at $r=0$ forbids the existence of regular stellar configurations. Hence, the Polyakov approximation must be modified at least in a central core of radius $r_{\rm core}\ll R$ around $r=0$, with $R$ being the radius of the star. A simple example is the following:
\begin{equation}\label{eq:4dimRSETCP}
F(r)=\frac{1}{r^2 + \alpha l_{\rm P}^2}.
\end{equation}
However, this   arbitrary choice does not guarantee    sensible results~\cite{Arrechea2021b}. We will work with functions $F(r)$ that differ from $1/r^{2}$ just for $r<r_{\rm core}$, including but not limiting ourselves to the specific example \eqref{eq:4dimRSETCP}. Our main result  is the existence of an entire family of possibilities for $F(r)$ characterized by the requirement of accommodating regular stellar configurations of arbitrary compactness.

\section*{Exterior solution}
For completeness, let us briefly summarize the properties of the external vacuum geometries corresponding to the solutions to Eqs. \eqref{Eq:SemiEinstein} in absence of a classical SET but with a nonzero $T_{\mu\nu}^{\rm ZP}$. These equations can be integrated from an asymptotically flat region inwards~\cite{Arrecheaetal2020}, obtaining the semiclassical counterpart of the Schwarzschild geometry.
Far from the gravitational radius, quantum deviations from the Schwarzschild spacetime amount to perturbative corrections to the mass and redshift functions. As the gravitational radius is approached, however, quantum corrections become non-perturbative and destroy the event horizon altogether.
The resulting geometry corresponds to an asymmetric wormhole with one asymptotically flat region and a neck lying above the gravitational radius. Deep inside the neck, the solution accommodates a null curvature singularity whose details are not needed here, as this geometry is just the external spacetime of a stellar-like object whose surface is located at a given radial position around the neck, either inside, outside, or at the neck itself.
From the selected surface inwards we integrate the Einstein equations with a classical source in addition to the always present semiclassical source.

\begin{figure}
    \centering
    \includegraphics[width=0.6\textwidth]{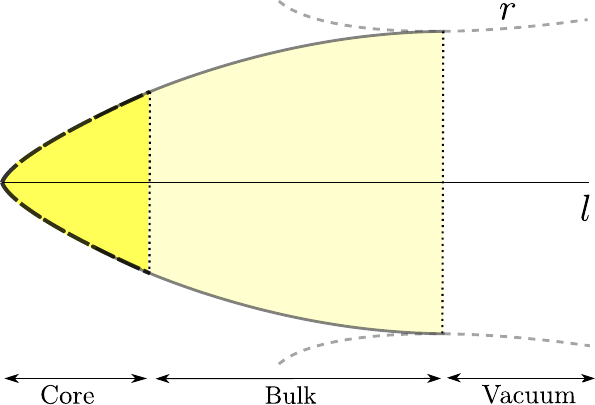}
    \caption{Pictorial representation of a semiclassical relativistic star. The areal radius $r$ of spheres is shown in terms of a proper coordinate $l$, defined as $dr/dl=\sqrt{1-C}$. The vacuum region (gray dashed lines) is the semiclassical Schwarzschild solution~\cite{Arrecheaetal2020} describing an asymmetric wormhole. The bulk (gray continuous lines) is well described by the Polyakov approximation. This approximation breaks down at the core of the star (black dashed lines), but it can be minimally modified to adequately describe this region.}
    \label{Fig:Stellar}
\end{figure}

\section*{Interior solution outside the core}
To complete the system of semiclassical field equations we assume an equation of state for the classical fluid, namely, uniform density
\begin{equation}\label{Eq:EoS}
    \rho(r)=\rho\equiv\text{const}.
\end{equation}
We integrate the semiclassical equations for this equation of state, for which the classical isotropic pressure $p$ adapts to the needs of the configuration. $\{C(r),p(r)\}$ then turn out to obey a system of two coupled differential equations, from which the function $\phi(r)$ is derived and thus, the entire geometry (see \cite{Arrechea2021b} for details).
The integration starts at the surface radius $R$, taking a compactness $C(R)=C_R$ and a pressure $p(R)=0$. One also has to decide the location of the star surface with respect to the putative wormhole neck. Let us first describe the properties of these geometries in the range $r \in (r_{\text{core}},R)$.

In~\cite{Arrechea2021b} we analyzed exhaustively these semiclassical equations of stellar equilibrium for the cutoff-regularized Polyakov approximation \eqref{eq:4dimRSETCP}  and obtained an entire catalog of regular and irregular solutions. The features of the solutions obtained in~\cite{Arrechea2021b} in the range $r \in (r_{\text{core}},R)$ are universal for all the choices of $F(r)$ considered here. For example, for compactness below but close to Buchdahl's $(C_R=8/9)$ we found regular stellar configurations perturbatively similar to their classical counterparts. Here, we report on a particular family of solutions which are found when the compactness of the star is close to the black hole limit, that is, amply surpassing the Buchdahl bound. In this limit, their qualitative form is not very much affected by the location of the surface with respect to the putative wormhole neck. These newly found stars display a three-layered structure that appears schematically depicted in Figure \ref{Fig:Stellar}. Having outlined previously the characteristics of the exterior (vacuum) solution, we now turn to describing the solution for the bulk.

\subsection*{Criticality and classical stellar solutions}
The inward integration takes as parameter the density $\rho$ \eqref{Eq:EoS}. In principle, given the initial conditions $\{R,C_R\}$, there is a \textit{critical} value $\rho_{\rm{c}}$ for which the configuration is regular all the way up to the center $r=0$. When this does not occur (as for stars with $C_R$ sufficiently close to $1$), we consider $\rho_c$ as the value of the density corresponding to a qualitative change in the behavior of the compactness (or the Misner-Sharp mass, equivalently) at the origin (see~\cite{Arrechea2021b} for a thorough discussion of this point). 

The various regimes in our numerical integrations for a star that surpasses the Buchdahl limit $C_R=8/9$ are represented in Figure~\ref{Fig:Separatrix}. For strictly classical stars $T^{\rm ZP}_{\mu\nu}=0$, the critical solution with $\rho=\rho_{\rm c}$ (thick line in  Fig. \ref{Fig:Separatrix}A) has vanishing Misner-Sharp mass at $r=0$, separating solutions with positive and negative mass at $r=0$.

Solutions with densities around $\rho_{\rm c}$ exhibit pressures that diverge at some radius far away from $r=0$ (thin dashed lines in Fig. \ref{Fig:Separatrix}A). Stars with $\rho\gg\rho_{\rm c}$ have this infinite pressure surface pushed inwards until, eventually, a solution where pressure diverges exactly at $r=0$ is reached (thick line in Fig. \ref{Fig:Separatrix}B). All solutions surrounding this separatrix between finite and infinite pressure solutions display a large negative mass at $r=0$. By increasing the value of $\rho$, pressure is made finite everywhere at the cost of making the compactness function singular at $r=0$.

\subsection*{Quasi-regular semiclassical stars}
\begin{figure}
    \centering
    \includegraphics[width=\textwidth]{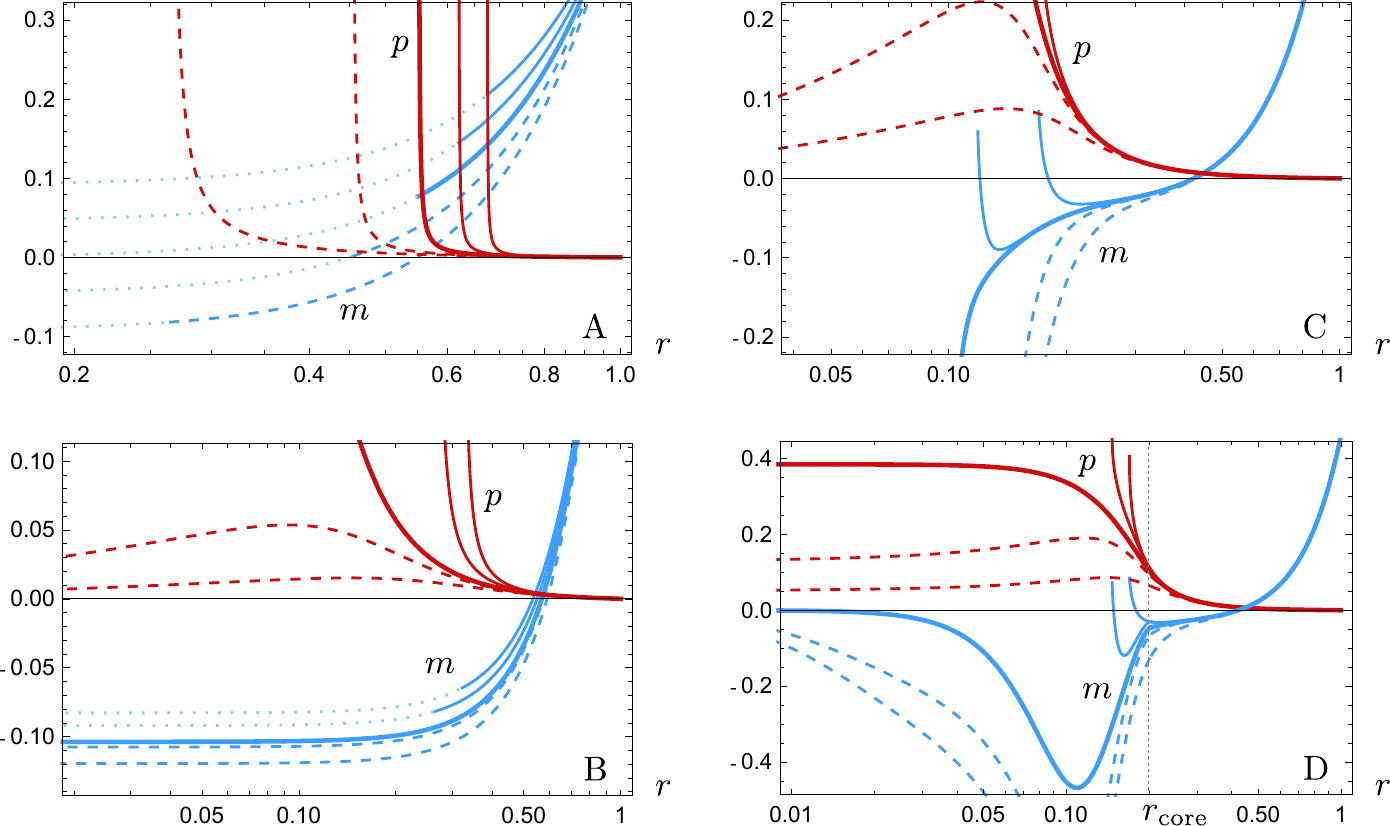}
    \caption{Plots of the pressure (red) and Misner-Sharp mass (blue) of solutions surrounding: (A) the classical critical solution with $\rho=\rho_{\rm c}$; (B) the classical separatrix in pressure; (C) the semiclassical critical solution with \mbox{$F(r)=1/\left(r^{2}+\alpha l_{\rm P}^{2}\right), ~\alpha=1.01$}; and (D) a semiclassical ultracompact stellar solution with a regular core of size \mbox{$r_{\text{core}}\simeq0.2$}. For every integration, we have chosen the values $R=1$ and $C_R=0.92$ for visualization purposes. The values of $\rho$ increase from right to left in the pressure profiles (e.g. thin dashed lines correspond to greater $\rho$ than thin continuous lines).}
    \label{Fig:Separatrix}
\end{figure}

The semiclassical situation obtained for the simple regulator choice \eqref{eq:4dimRSETCP} shown in Fig. \ref{Fig:Separatrix}C is rather different from the classical scenario and already reveals appealing properties. The first one is that the critical solution for $\rho=\rho_{\rm c}$ represents two separatrices that appear {\em together} in mass and pressure (the separatrix corresponds again to the thick lines in Fig. \ref{Fig:Separatrix}C).
The second compelling property is manifested for solutions where $\rho \lesssim \rho_c$. For these sub-critical stars close to criticality (thin continuous lines in Fig. \ref{Fig:Separatrix}C), $m(r)$ acquires negative values followed by a bounce back to positive values, eventually reaching a surface where a wormhole neck is finally formed; the pressure reaches the neck with finite values (we have stopped these integrations at the neck since it is a singular surface for the $r$ coordinate). There are sub-critical solutions with arbitrarily small necks.
 
Both aforementioned characteristics, i.e. a simultaneous separatrix behaviour in mass and pressure and a vanishing central mass, must be fulfilled by any regular solution. Therefore, we state that the Polyakov approximation manages to generate quasi-regular ultracompact stars, in the following sense. For the geometry to be regular at the center of spherical symmetry, the mass function must vanish there while having a finite pressure. Hence, if we consider a small core around the center, regular configurations will have small masses (due to continuity) and finite pressures. This is not the case in the classical theory (Figs. \ref{Fig:Separatrix}A and \ref{Fig:Separatrix}B), where finite pressures for configurations beyond the Buchdahl limit require large negative values of the mass~\cite{Arrechea2021b}. Crucially, semiclassical physics is able to produce ultracompact configurations compatible with this vanishing of the mass and a finite pressure at a central core.
Remarkably, the core can be Planck-sized for an ultracompact stellar object of say radius $R\sim3~{\rm km}$. Strict regularity is not fulfilled because of the singularity at $r=0$ of the Polyakov approximation \eqref{eq:4dimRSET}. The cutoff-regularized Polyakov approximation \eqref{eq:4dimRSETCP} also fails to provide a strictly regular geometry (Fig. \ref{Fig:Separatrix}C), as a singularity beyond a wormhole neck is produced inside the core.

\section*{Core regularization}

The Polyakov approximation fails to capture the correct physics close to the radial origin. This observation prompts the following questions: How off the mark is this approximation? Is it possible that slight deformations  lead to regular stellar solutions? This is the motivation behind the family of approximations studied here, characterized by functions $F(r)$ that differ from $1/r^{2}$ just for $r<r_{\rm core}$. It is important to stress that deforming the Polyakov approximation is not a choice, but a necessity to avoid its singular nature at $r=0$. We consider the minimal extensions to achieve this goal. We now show that these minimal deformations of the Polyakov approximation suffice to produce regular configurations in a generic way, i.e., there exists a whole family of functions $F(r)$ leading to regular configurations.

We follow a reverse-engineering logic which consists in making an ansatz for a  regular geometry  in the range $r \in (0,r_{\text{core}})$ and then obtaining the regulator $F(r)$ that sources the geometry via the RSET, in case it exists. We derive an expression for $C$ from the $rr$ component of the semiclassical equations \eqref{Eq:SemiEinstein} and replace it in the $tt$ component. Furthermore, through conservation of the classical SET, we find the relation
\begin{align}
    p''&=
        \mathcal{D}\left[\mathcal{A}_{0}+\mathcal{A}_{1}\left(p'\right)+\mathcal{A}_{2}\left(p\right)'^{2}+\mathcal{A}_{3}\left(p'\right)^{3}\right],
    \label{Eq:Pres}
\end{align}
where 
\begin{align}
    \mathcal{A}_{0}=
    &
    -8\pi r\left(\rho+p\right)^{3}\left(\rho+3p\right),\nonumber\\
    \mathcal{A}_{1}=
    &
    ~4\left(\rho+3p\right)^{2}\left[6\pi r^{2}\left(\rho+p\right)+4\pi F l_{\rm P}^{2}r^{2}p -1\right],\nonumber\\
    \mathcal{A}_{2}=
    &
    -r\left(\rho+p\right)\left[16\pi r^{2}\left(\rho-2p\right)-l_{\rm P}^{2}\left(2F+rF'\right)\right.\nonumber\\
    &
    \left.\hspace{1.8cm}+8\pi Fl_{\rm P}^{2} r^{2}\left(\rho+5p\right)-8\pi F'l_{\rm P}^{2}r^{3}p-6\right],\nonumber\\
    \mathcal{A}_{3}=
    &
    ~F l_{\rm P}^{2}r^{2}\left[8\pi r^{2}\left(\rho-p\right)-l_{\rm P}^{2}\left(2F-rF'\right)\right.\nonumber\\
    &
    \hspace{1.3cm}\left.-8\pi F'l_{\rm P}^{2}r^{3}p-2\right],\nonumber\\
    \mathcal{D}=
    &
    ~2r\left(1-l_{\rm P}^{2}F\right)\left(\rho+p\right)^{2}\left(1+8\pi r^{2}p\right).
\end{align}
By imposing an ansatz for the pressure and its derivatives, Eq.~\eqref{Eq:Pres} becomes a first-order differential equation for the regulator $F$ which, upon solving, determines the entire core geometry. Naturally, if the resulting $F$ is everywhere regular inside the core, the star spacetime metric will be regular as well.

We consider a pressure profile for the core (whose classical energy density is constant according to \eqref{Eq:EoS}) that is regular and has a global maximum at $r=0$. At $r_{\text{core}}$, continuity of the metric enforces pressure to be continuous up to its second derivative.
The simplest function that satisfies these conditions is a fifth-order polynomial
\begin{equation}
p= p_{0}+p_{0}''r^{2}/2+c_{0} r^3+c_{1}r^4+c_{2}r^5,
\end{equation}
where the pressure at the origin $p_{0}$ and its second derivative $p_{0}''$ are positive and negative constants, respectively. Determining the coefficients $\left\{c_{i}\right\}_{i=0}^{2}$ is straightforward given the above conditions. Now, taking a fixed solution for the bulk region $r\in(r_{\text{core}},R)$ (hence, a particular pressure profile) and a core size $r_{\text{core}}$, the pressure function inside the core is determined upon fixing the two remaining free parameters $\{p_{0},p_{0}''\}$.

\begin{figure}
    \centering
    \includegraphics[width=0.7\columnwidth]{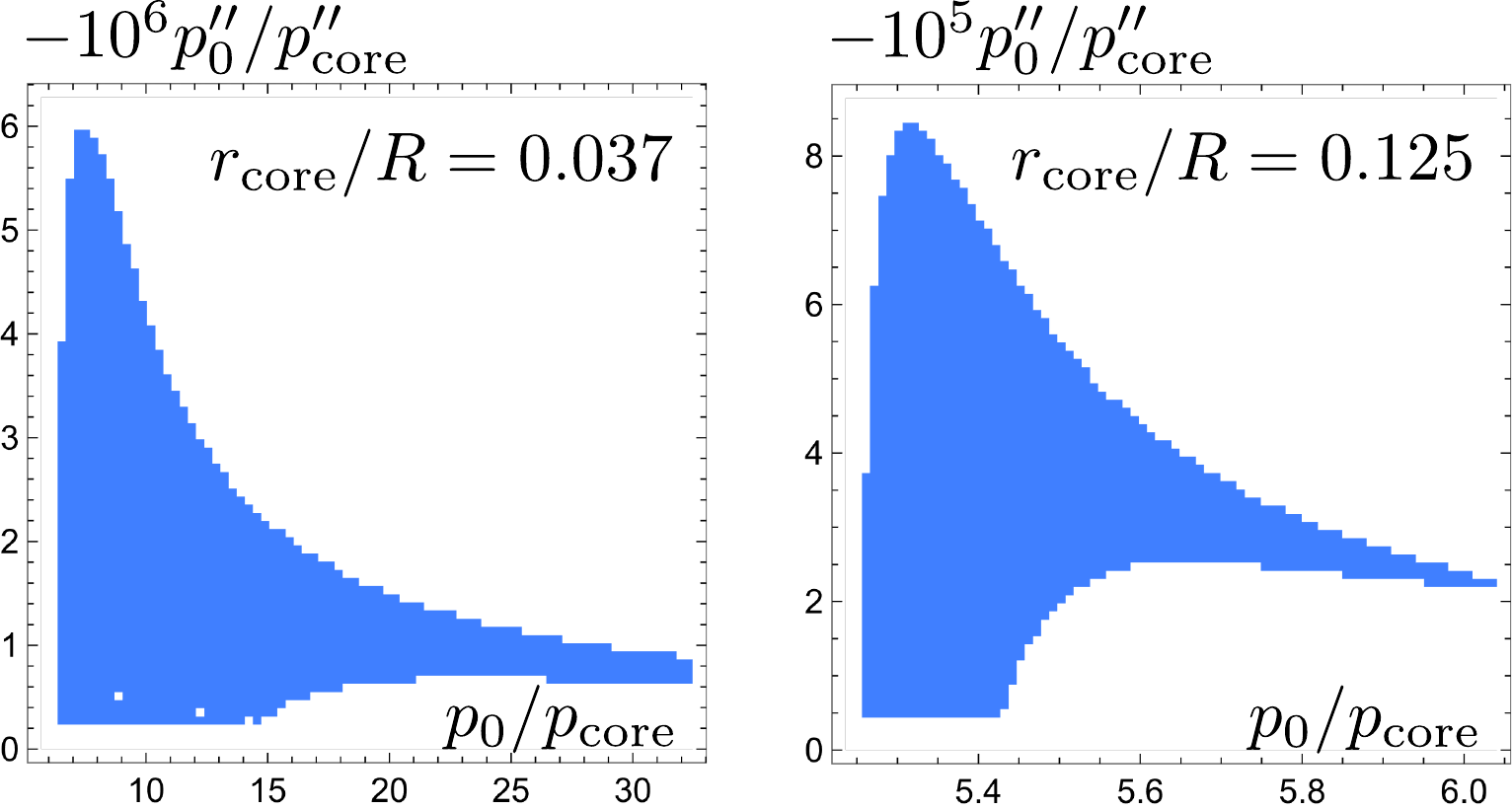}
    \caption{
    The left and right panels denote the parameter space for the analytical-pressure core of stars with $R\simeq 163,~C_R=1-10^{-4}$. We have chosen $r_{\text{core}}/R=0.037$ and $r_{\text{core}}/R=0.125$ for the core radius, respectively. The colored region denotes entirely regular solutions. The horizontal axis is the quotient between the central and the core-boundary pressures $p_{0}$, $p_{\text{core}}$. The vertical axis is the quotient between the second derivatives of the pressure at the origin, $p''_{0}$, and at the core boundary, $p''_{\text{core}}$.}
    \label{Fig:PhaseDiagram}
\end{figure}

We have performed a numerical exploration of a wide range of values of the parameters $\{p_{0},p_{0}''\}$ given a set of fixed solutions for $r>r_{\text{core}}$. The results are represented in Fig. \ref{Fig:PhaseDiagram}, where whole parametric regions of regular solutions are shown. These regions exist for central cores of any size.

This result is remarkably non-trivial, as it is not guaranteed that a prescribed geometry will be compatible with  the Polyakov RSET multiplied by a function. It might have happened that no $F(r)$ existed for any regular ansatz. Hence, that this compatibility is realized for the simple polynomial example described above is a strong indication that the Polyakov approximation is able to capture an important fraction of the relevant physics.

The strictly regular solutions we have obtained have a clear interpretation in terms of the regulating functions $F$. By modifying the regulator inside the core we are distorting the space of solutions (Fig. \ref{Fig:Separatrix}D) so that the new critical solution corresponds to a regular configuration. The regular separatrix solution exhibits an interior region of negative mass (encompassing the core and part of the bulk) that exerts the gravitational repulsion necessary to sustain the whole structure for values of $C_R$ for which a sphere composed of a classical fluid alone would inevitably collapse under its own gravity.

\begin{figure}
    \centering
    \includegraphics[width=0.6\textwidth]{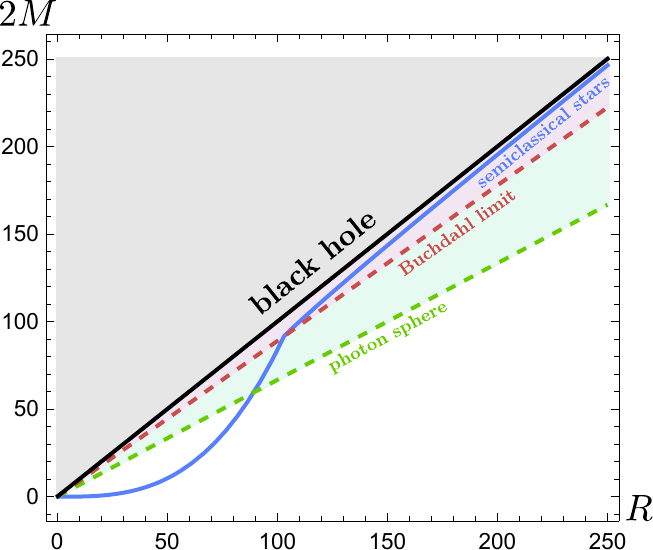}
    \caption{Mass-radius diagram of semiclassical relativistic stars with $\rho=10^{-5}$. The black line represents the compactness parameter of black holes $(C_R=1)$, the dashed red line denotes the Buchdahl compactness bound $(C_R=8/9)$, and the dashed green line is the minimum compactness of objects that exhibit a photon sphere $(C_R=1/3)$. The blue curve represents semiclassical relativistic stars. For stars surpassing the Buchdahl limit, the total mass $M$ grows approximately linearly with the radius $R$. Each point within the blue curve admits entire families of regulator functions $F$ for which the whole geometry is regular.}
    \label{Fig:MassRadius}
\end{figure}

As the bulk geometry of the star is unaffected by the characteristics of the regular core, the surface compactness $C_R$ and radius $R$ are fixed constants in our integrations. Therefore, the mass-radius diagram for semiclassical relativistic stars shown in Fig. \ref{Fig:MassRadius} is independent of $r_{\text{core}}$. The most remarkable feature that we can extract from this diagram is that semiclassical relativistic stars can be arbitrarily close to the black hole limit. 

\section*{Parametrized shapes}

To end this manuscript we provide a family of analytic geometries that shows the main characteristics of semiclassical relativistic stars. They exhibit a negative mass region in the interior of the structure together with a redshift   that decreases inwards.
This family  accommodates qualitatively to foreseeable solutions found using different approximation schemes to the RSET. Alternatively, it can be taken as a parametrized phenomenological approach to this type of ultracompact objects. 

Defining $\hat{r}=r/R$, our 5-parameter family of metrics is:
\begin{align}
    ds^{2}= -e^{2\phi(\hat{r})} dt^2 +R^{2}\left[1-C(\hat{r})\right]^{-1}d\hat{r}^{2} + R^{2}\hat{r}^{2} d \Omega_2^2,
   \label{Eq:BSMetric}
\end{align}
with 
\begin{align}\label{Eq:ParFamily}
& e^{2\phi(\hat{r})}=
    \begin{cases}
    1- {C_R /\left( \hat{r} + \beta_{0} \right)},
    &  1 \leq \hat{r} < \infty 
    \\
    \beta_{1}a_{0}^{\beta_{2}\hat{r}^{2}} +a_{1}\hat{r}^{6}e^{a_{2}\left(\hat{r}-1\right)},
    & 0  \leq \hat{r} < 1
    \end{cases}
\nonumber \\
& C(\hat{r})=
    \begin{cases}
    C_{R}/\hat{r},
    &  1 \leq \hat{r} < \infty \\
    \beta_{3}\left[\cos\left(\beta_{4} \hat{r}\right)-1\right]e^{-\beta_{4}\hat{r}} +a_{3}\hat{r}^{2},
    & 0  \leq \hat{r} < 1
    \end{cases}
\end{align}
The constants $\left\{a_{i}\right\}_{i=0}^{3}$   depend on the 5 independent parameters $\left\{\beta_{i}\right\}_{i=0}^{4}$: 
\begin{align}
    a_{0}=
    &
    \frac{1-C_R+\beta_{0}}{\beta_1\left(1+\beta_{0}\right)},\quad a_{1}=
    1-\frac{C_R}{1+\beta_{0}}-\beta_{1}a_{0}^{\beta_2},\nonumber\\
    a_{2}=
    &
    \left\{2\beta_{1}a_{0}^{\beta_{2}}\left[3-2\beta_{2}\log a_{0}\right]+\frac{C_R\left(7+6\beta_{0}\right)}{\left(1+\beta_{0}\right)^{2}}-6\right\}a_{1}^{-1},\nonumber\\
    a_{3}=
    &
    \beta_{3}e^{-\beta_{4}}\left(1-\cos \beta_{4}\right)+C_{R}.
\end{align}

\begin{figure}
    \centering
    \includegraphics[width=0.6\textwidth]{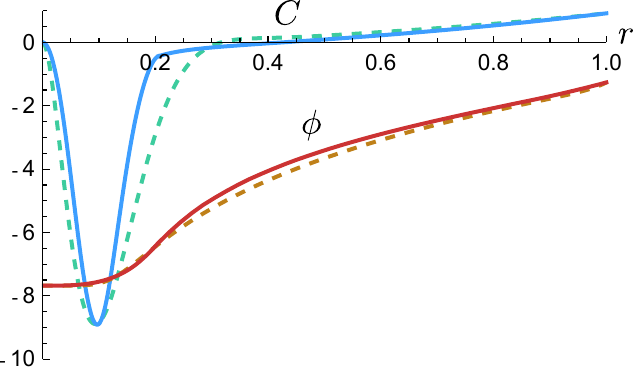}
    \caption{Comparison between a numerical solution describing a semiclassical ultracompact object (continuous lines, compactness $C(r)$ in blue and redshift $		\phi(r)$ in red) and a particular solution of the parametrized family of geometries \eqref{Eq:ParFamily} (dashed lines). The numerical solution is the separatrix from Figure \ref{Fig:Separatrix}D.
     The parameters of the analytical fit are $\beta_{0}=10^{-4},~\beta_{1}=2.14\times 10^{-7},~\beta_{2}=0.905,~\beta_{3}=42.8$ and $\beta_{4}=16.8$.}
    \label{Fig:ParamShapes}
\end{figure}

Thus, the family of metrics \eqref{Eq:ParFamily} is characterized by five
form parameters $\{\beta_i\}_{i=0}^4$  with a clear physical interpretation. $\beta_{0}$ introduces an offset in the redshift $e^{2\phi(r)}$ associated with the fact that the external semiclassical metrics we have found are almost Schwarzschild up to very close to the gravitational radius.   $\beta_1\in (0,e^{2\phi(R)})$ represents the redshift at the origin. $\beta_2\in\left[0,1\right)$ controls the flatness of the redshift profile in the interior region. Finally $\beta_3,\beta_4>0$ determine the width and depth of the negative energy internal region. Figure \ref{Fig:ParamShapes} shows a comparison between a numerical solution and the corresponding analytical fit (see the supplemental material \cite{Supplemental} to visualize how shifting the parameters modifies the solution).

\section*{Conclusion}
Neutron stars are the most compact relativistic stars known to exist. We have shown that semiclassical gravity can accommodate more compact stellar configurations supported by  quantum vacuum polarization. This result has been obtained using a well-motivated semiclassical source given by a minimal deformation of the Polyakov approximation. A clear extension of this work is to analyze whether similar solutions exist when using more refined proposals for the RSET in (3+1) dimensions~(e.g.~\cite{Andersonetal1995}). We have found preliminary evidence that the existence of solutions as described here persist in more elaborate approximations, and will present the corresponding results elsewhere.
These investigations open the possibility for the existence of new stages of stellar evolution beyond relativistic stars.

\subsection*{Acknowledgments}
The authors thank Gerardo Garc\'{\i}a-Moreno and Valentin Boyanov for very useful discussions and the anonymous reviewers for their comments. Financial support was provided by the Spanish Government through the projects PID2020-118159GB-C43/AEI/10.13039/501100011033, PID2020-118159GB-C44/AEI/10.13039/501100011033, PID2019-107847RB-C44/AEI/10.13039/501100011033, and by the Junta de Andalucía through the project FQM219. This research is supported by a research grant (29405) from VILLUM fonden. CB and JA acknowledges financial support from the State Agency for Research of the Spanish MCIU through the “Center of Excellence Severo Ochoa” award to the Instituto de Astrofísica de Andalucía (SEV-2017-0709).

\bibliographystyle{unsrt}
\bibliography{biblio-semiclassical}

\end{document}